\documentclass[aps,pra,twocolumn,groupedaddress,showpacs,superscriptaddress]{revtex4-1}
\usepackage{graphicx,amsmath}
\usepackage{amssymb}

\begin{document}
\title{Hubbard model on Semiclassical approximation in combination with an optimizer based on GPU technology}
\author{Hayun Park}
\affiliation{Department of Liberal Studies, Kangwon National University, Samcheok, 25913, Republic of Korea}
\author{Hunpyo Lee}
\affiliation{Department of Liberal Studies, Kangwon National University, Samcheok, 25913, Republic of Korea}
\email{Email: hplee@kangwon.ac.kr}
\date{\today}

\begin{abstract}
We developed a semiclassical approximation method in combination with an adaptive moment estimation 
optimizer (SCA + ADAM) approach based on the PyTorch plus CUDA library on a the graphics processing unit 
(GPU). This method was employed to evaluate one-particle properties of the Hubbard model with long-range 
spatial correlations within an appropriate computing duration. The method was applied to the ionic Hubbard 
model on a two-dimensional square lattice with long-range spatial correlations. The computation time was 
evaluated as a function of the lattice size on the central processing unit and GPU. Herein, we also 
discuss the density of states and antiferromagnetic (AF) order parameter in the Hubbard model without the 
ionic potential and compare the results with those of the Hartree-Fock approximation. Finally, we present 
the one-particle properties and order parameter in charge density wave, AF metal and AF insulator of the 
ionic Hubbard model.
\end{abstract}

% 71.10.Fd    Lattice fermion models (Hubbard model, etc.)
% 71.27.+a    Strongly correlated electron systems; heavy fermions
% 71.30.+h    Metal-insulator transitions and other electronic transitions
\maketitle

\section{Introduction\label{Introduction}}

Development of optimizers based on the graphics processing unit (GPU) technology has resulted in several 
breakthroughs in the field of artificial neural network. In addition, such optimizers have been 
extensively applied in various fields of computational science and economics involving optimization 
problems. Therefore, assessing the applicability of an optimizer to a GPU to solve interesting physical 
problems has garnered considerable attention in recent years.

The Hubbard model, which describes the competition between the kinetic energy of electrons and repulsive 
Coulomb potential energy of electrons, is one of the most popular and fundamental problems in 
physics~\cite{Imada1998}. The solutions of the Hubbard model are anticipated to reveal the origin of the 
high-temperature cuprate superconductivity, unconventional insulating behavior, and non-Fermi liquids 
appearing in two-dimensional (2D) electronic systems. Deriving an exact solution in the thermodynamic 
limit using the exact 
diagonalization (ED) method is limited, because the size of the Hamiltonian increases exponentially with 
an increase in the number of sites in a lattice~\cite{Ohta1994,Go2017}. The unbiased quantum Monte Carlo 
(QMC) approach exhibits the infamous sign problem in the repulsive Fermionic Hubbard 
model~\cite{Hirsch1986,Gull2011}. Moreover, the QMC method is restricted to moderately sized lattices owing to its 
computational burden. Therefore, despite numerous numerical efforts, development of novel numerical and 
theoretical methods for solving the Hubbard model continues to remain at the fore of research in this 
field~\cite{Georges1996,Evers2008,Moukouri2001,Kyung2003,Maier2005}.

A semiclassical approximation (SCA) approach is a well-established method for solving the Hubbard model~\cite{Okamoto2005,Lee2013}. 
In the SCA, the onsite repulsive Hubbard interaction is decoupled into charge and spin fluctuations. A 
potential function with auxiliary fields is created via continuous Hubbard-Stratonovich transformation of 
the spin fluctuation term, and charge fluctuation is ignored. The number of auxiliary variables is equal 
to one of the sites on the lattice and increases with the increasing number of sites on the 
lattice. Determining the variables that minimize the potential energy is necessary in the SCA approach. 
However, the computational cost increases polynomially with an increase in the number of sites, thus impeding exploration of the Hubbard model with long-range spatial 
correlations. Accordingly, to date, the SCA has only been used in combination with the cluster dynamical 
mean field theory approach for a small-sized lattice system.

\begin{figure}
\includegraphics[width=0.95\columnwidth]{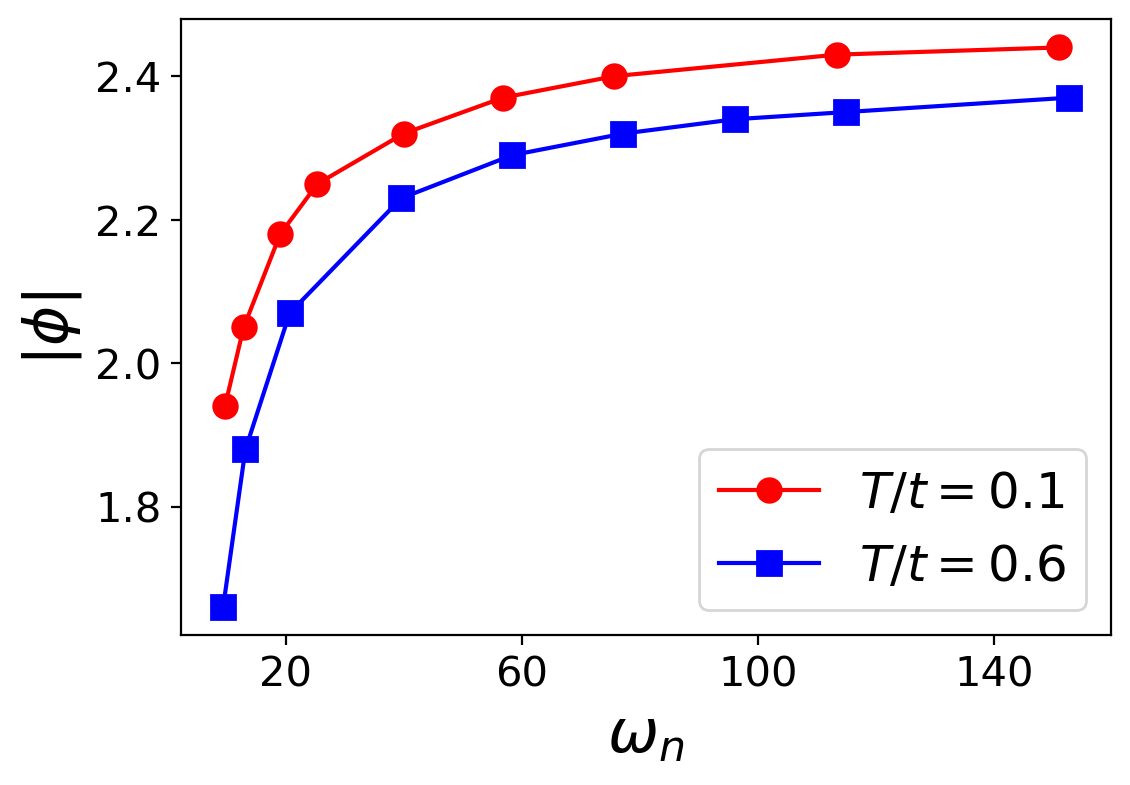}
\caption {\label{Fig1} (Color online) Antiferromagnetic order parameter $\vert \phi \vert$ as a function
Matsubara frequency $\omega_n$ in the Hubbard model on half-filled two-dimension (2D) $20 \times 20$ 
square lattice at Hubbard interaction $U/t=6.0$ for temperature $T/t=0.1$ and $0.6$. $\omega_n$ is given 
as $\omega_n = (2n + 1) \pi T$, where $n$ are integer numbers.}
\end{figure}

In this study, we developed an SCA integrated adaptive moment estimation optimizer (SCA+ADAM) approach 
based on the PyTorch plus CUDA library on a GPU. The auxiliary variables of the potential function created 
in the SCA were rapidly determined using a parallelized auto-gradient approach in the ADAM optimizer on 
the GPU~\cite{Jimmy2014}. Thus, the proposed integrated SCA+ADAM approach can be applied to a large Hubbard model with 
long-range spatial correlations. To evaluate the effectiveness of our SCA+ADAM approach, the approach was 
applied on the ionic Hubbard model of a half-filled 2D $L \times L$ square lattice, wherein electronic 
hopping, onsite periodic potential, and onsite repulsive Coulomb interactions induced a metallic state, 
band insulator (BI), and antiferromagnetic (AF) insulator, respectively. First, we examined the computational costs for various parameters on the central processing unit (CPU) and GPU. Next, we evaluated the 
density of states $\rho_{\sigma} (\omega)$ and AF order parameter $\phi$ of the 
Hubbard model without an onsite periodic potential. We also compared our SCA+ADAM results with those 
obtained using the Hartree-Fock method. Finally, the physical properties, phases and AF order parameter of 
the ionic Hubbard model were investigated.

The remainder of this paper is organized as follows: Section~\ref{model} describes the Hamiltonian of the 
ionic Hubbard model and formalism of the SCA approach. In Section~\ref{Result}, computational
cost and the results of $\rho_{\sigma}(\omega)$ and $\phi$ for both the pure and ionic Hubbard 
models are presented. Finally, the major conclusions drawn from the findings of this study are presented 
in Section~\ref{summary}.

\section{Hamiltonian and Formalism of semiclassical approximation\label{model}}

In this study, we considered the ionic Hubbard model of a half-filled 2D $L \times L$ square 
lattice~\cite{Bouadim2007,Go2011}. Since the discovery of high-temperature cuprate superconductors, 
several studies have been conducted on 
the 2D Hubbard model of a square lattice. In addition, the results of the Hubbard model of a 2D $L \times 
L$ square lattice with half-filled particle-hole symmetry are well understood. Therefore, we believe that 
this model is a useful benchmark for examining novel method.

\begin{figure}
\includegraphics[width=0.95\columnwidth]{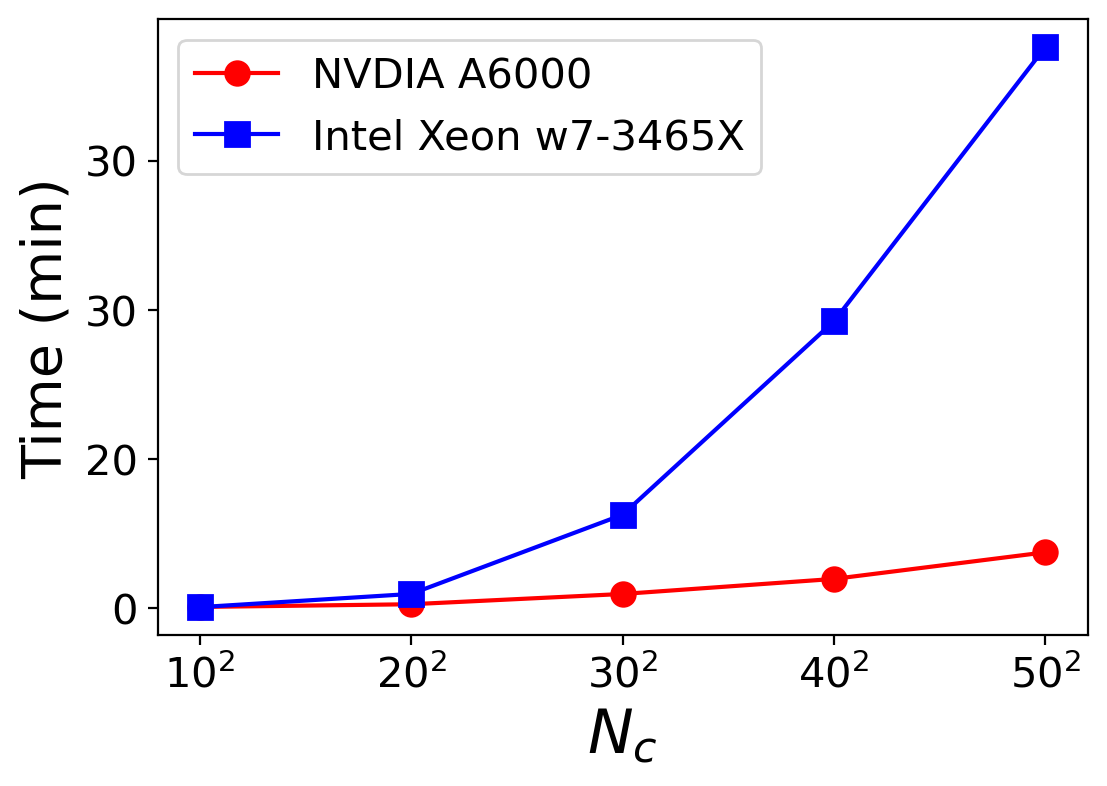}
\caption {\label{Fig2} (Color online) Computing time (min) as a function of $N_c$ on CPU (Intel Xeon 
w7-3465X) and GPU (single NVIDIA RTX-6000A) to determine optimal $\vert \phi \vert$ in the Hubbard model 
on 2D $L \times L$ square lattice. $N_c$ is defined as $N_c=L^2$.}
\end{figure}

The Hamiltonian of the ionic Hubbard model is expressed as
\begin{eqnarray}
H=-t\sum_{i\epsilon A,j\epsilon B,\sigma}(c_{i\sigma}^{\dag}c_{j\sigma}+\text{h.c.})
-\Delta \sum_{i\epsilon A}{n_i}+\Delta \sum_{i\epsilon B}{n_i} \nonumber \\
+U\sum_{i}n_{i\uparrow}n_{i\downarrow}
-\mu \sum_{i\sigma}n_{i},
\label{IHH}
\end{eqnarray}
where $t$, $\mu$ and $U$ are the nearest-neighbor hopping, chemical potential and repulsive Coulomb 
interactions, respectively. Here, $c_{i\sigma}^{\dagger}$ and $c_{i\sigma}$ are the electron creation and 
annihilation operators at site $i$ with spin $\sigma$, respectively. $\Delta$ is the ionic staggered 
potential that alternates sign between sites in sublattices $A$ or $B$. In this study we considered only 
the half-filled case. The energy scale $t$ was set to $t=1$, and the size $N_c$ of the noninteracting 
Hamiltonian with $U/t=0$ was $N_c=L^2$.

Here, we present the formalism of the SCA approach. The partition function of the Hubbard model is 
expressed as
\begin{equation}
Z = \int D[c^{\dagger},c] \exp[-S_{\text{eff}}].
\end{equation}
Here, the effective action can be written as
\begin{equation}
S_{\text{eff}} = \int_{0}^{\beta} d\tau \int_{0}^{\beta} d\tau' c^{\dagger}
(\tau) \hat{a}(\tau - \tau')c(\tau') + U\int_{0}^{\beta} d\tau n_{\uparrow}
(\tau) n_{\downarrow} (\tau),
\end{equation}
where $c^{\dagger}$ and $c$ are the Grassmann variables, 
$c^{\dagger}=(c_{\uparrow}^{\dagger},c_{\downarrow}^{\dagger})^{\text{T}}$
and $c=(c_{\uparrow},c_{\downarrow})^{\text{T}}$. $\hat{a}(\tau - \tau')$ denotes the inversion matrix of 
$2N_c \times 2N_c$ Hamiltonian. $\beta$ denotes the inverse temperature $T$. $n_{\uparrow}(\tau) 
n_{\downarrow} (\tau)$ are transformed into
\begin{equation}
n_{\uparrow}(\tau) n_{\downarrow} (\tau)=\frac{1}{4}(N(\tau)^2 - M(\tau)^2),
\end{equation}
where $N(\tau) =n_{\uparrow} (\tau)+n_{\downarrow} (\tau)$ and $M(\tau) = n_{\uparrow}(\tau)-
n_{\downarrow}(\tau)$ mean the charge and spin fluctuations, respectively. The SCA approach ignores the 
charge fluctuations $N(\tau)$.

\begin{figure}
\includegraphics[width=0.95\columnwidth]{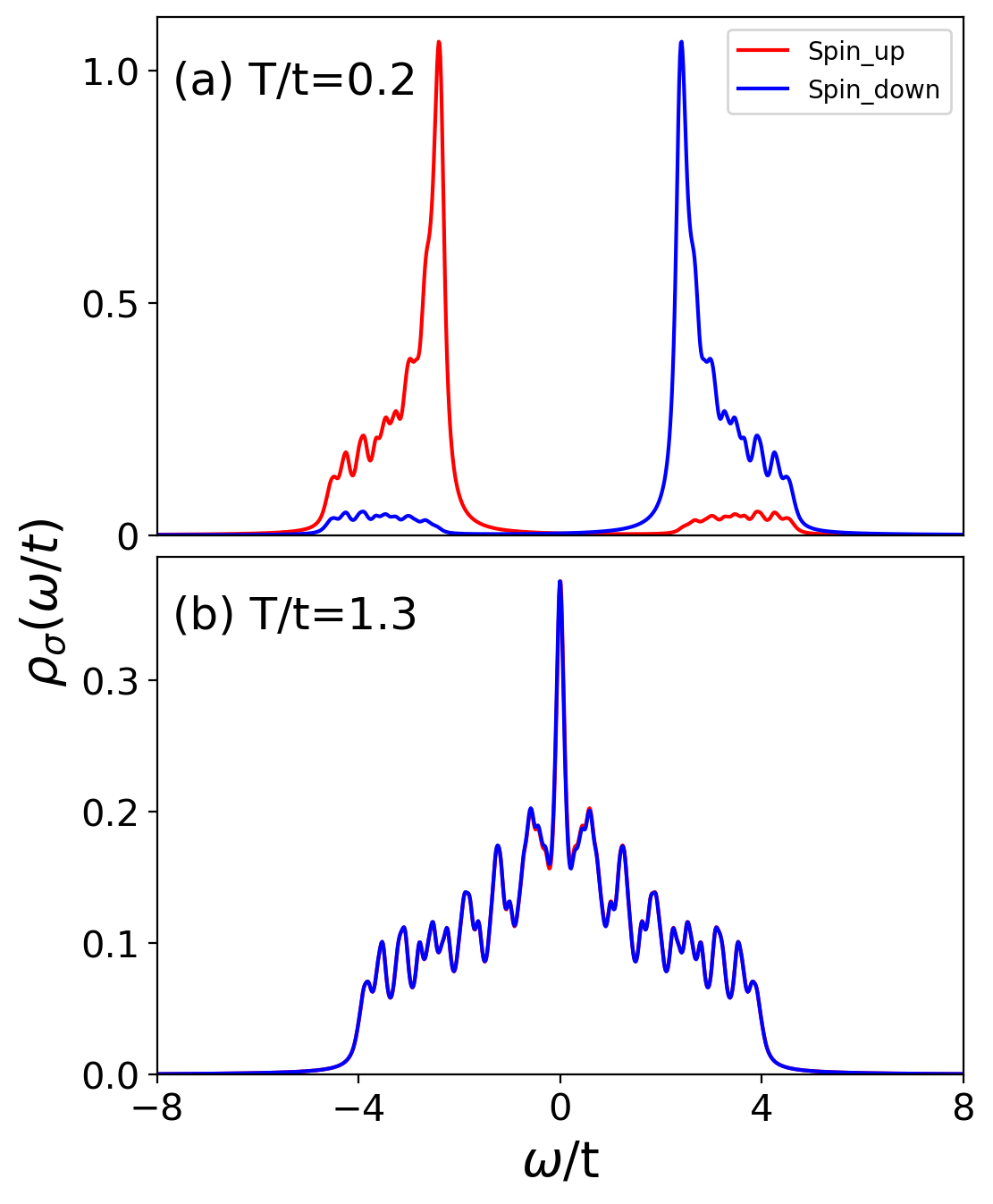}
\caption {\label{Fig3} (Color online) Density of states $\rho_{\sigma} (\omega/t)$ as a function of real
frequency $\omega/t$ at spin $\sigma$ in the Hubbard model on half-filled 2D $20 \times 20$ square lattice at $U/
t=6.0$ for (a) $T/t=0.2$ and (b) $1.3$.}
\end{figure}

The Gaussian form of the approximated partition function is derived via the Hubbard-Stratonovich 
transformation of the spin fluctuation term $M(\tau)$. The partition function can be rewritten as
\begin{equation}\label{Partition}
Z = \int \int D[c^{\dagger}c] \int_{-\infty}^{\infty} d\phi e^{-H(\phi)},
\end{equation}
where $H(\phi) = -\frac{\beta \phi^2}{4U} + \int d\tau \int d\tau' c^{\dagger}(\tau) (\hat{a}(\tau
\tau') + \frac{1}{2} \phi \sigma_z \delta(\tau - \tau')) c (\tau')$. $\sigma_z$ is the Pauli matrix of 
$z$-component. All $\tau$-dependent auxiliary fields $\phi_j (\tau)$ are approximated into static $\phi_j 
(\tau)=\phi_j$ without $\tau$-dependence. After Fourier transformation and Grassmann integration of the 
partition function given in Eq.~(\ref{Partition}), the final partition function of the SCA is given by
\begin{equation}\label{SCA}
Z = \int_{-\infty}^{\infty} d\phi_1 \ldots \phi_{N_c} e^{-\beta V(\phi)},
\end{equation}
where $V(\phi)$ denotes the potential generated by SCA. The detailed $V(\phi)$ is expressed as follows:
\begin{equation}\label{Eq2}
V(\phi)= \frac{1}{4U}\sum_j^{N_c} \phi_j^2 - T \sum_{\omega_n} \ln \det [-\beta (\hat{a} (\omega_n) +
\wedge (\phi)],
\end{equation}
where $\hat{a} (\omega_n)$ and $\wedge (\phi)=\text{diag} (\phi_1\sigma_z,\ldots,\phi_{N}\sigma_z)$
are $2N_c \times 2N_c$ matrices. The Matsubara frequencies $\omega_n$ are represented by
$\omega_n = (2n + 1) \pi T$. The $\phi$ values with the lowest possible energy of $V(\phi)$ were computed 
using an ADAM optimizer based on the PyTorch library. Here, $\phi$ is the AF order parameter in SCA 
approach. After 
determining all $\phi$, we modify $H$ into an approximated $2N_c \times 2N_c$ Hamiltonian $H_{\text{SCA}}$
with $\phi$. $\rho(\omega)$ is determined as follows:
\begin{equation}
\rho_{\sigma} (\omega) = -\frac{1}{\pi}\rm{Im}(\frac{1}{\omega + i \eta + \mu - H_{\text{SCA}}}),
\end{equation}
where $\eta$ denotes the broadening factor.

\section{Result\label{Result}}

First, we discuss the computation of Eq.~(\ref{Eq2}) to confirm a feasible computation time. Computing 
the determinant of the $2N_c \times 2N_c$ matrix in Eq.~(\ref{Eq2}) requires iterations of infinite 
$\omega_n$ values, which is not feasible. However, the high-frequency parts of $\omega_n$ are less 
important for determining the exact AF order parameter $\vert \phi \vert$ within the approximation. Thus, 
we computed $\vert \phi \vert$ as a function of $\omega_{n_{\text{max}}}$; here, $\omega_{n_{\text{max}}}$ 
is the maximum Matsubara value considered in the determinant computation in Eq.~(\ref{Eq2}). Fig.~\ref{Fig1} shows 
AF order parameter $\vert \phi \vert$ as a function of $\omega_{n_{\text{max}}}$ for the Hubbard model 
with $T/t=0.1$ and $0.6$. $\vert \phi \vert$ increases with increasing $\omega_{n_{\text{max}}}$ and nearly converges 
at $\omega_{n_{\text{max}}}=100$ in both the cases. Although $\phi$ is related to the gap size in the AF 
insulator, it does not affect the physical phase. We believe that it is sufficient to set 
$\omega_{n_{\text{max}}}=100$ to reduce the computational burden on the results shown in Fig.~\ref{Fig1}.

\begin{figure}
\includegraphics[width=0.95\columnwidth]{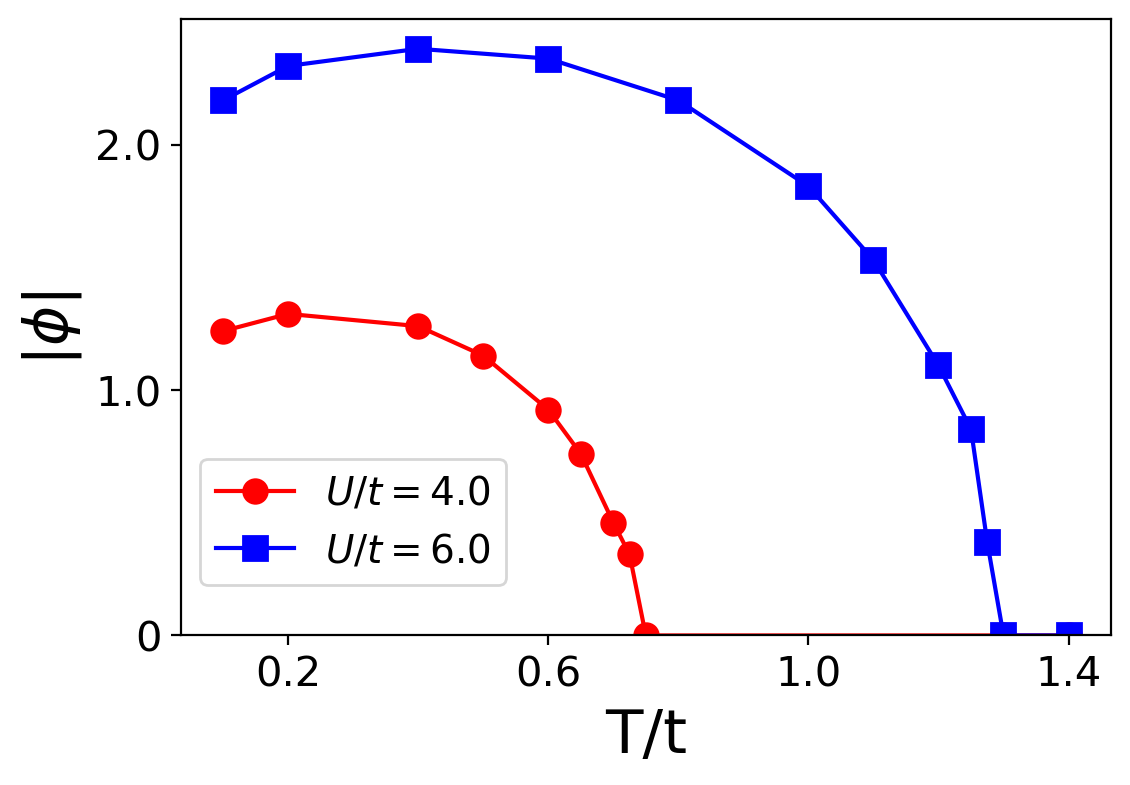}
\caption {\label{Fig4} (Color online) $\vert \phi \vert$ as a function of $T/t$ in the Hubbard model on 
half-filled 2D $20 \times 20$ square lattice for $U/t=4.0$ and $6.0$.}
\end{figure}

Next, we discuss the computational expenses and limitations in the system size for the CPU and GPU. The 
computation of the determinant of the $2N_c \times 2N_c$ matrix via iteration of $\omega_{n}$ and the 
optimization to determine $\vert \phi \vert$, which has the lowest possible energy in $V(\phi)$ 
in Eq.~(\ref{Eq2}), are the most computationally intensive steps; here, Eq.~(\ref{Eq2}) includes loops, 
matrix determinant calculations, and optimizations of potential. The program was solved using the ADAM 
optimizer with an auto-gradient approach based on the Pytorch library. Fig.~\ref{Fig2} shows the 
computation time as a function of $N_c$ (associated hardware: an Intel Xeon w7-3465X CPU and a single 
NVIDIA RTX-A6000 GPU). Because the GPU first frees up memory space for parallel computation, a single GPU 
can count up to $N_c=60^2$ for Matsubara frequency $n=15$ in Eq.~(\ref{Eq2}) with slowly increasing 
computing time, as shown in Fig.~\ref{Fig2}. However, the computational costs of the CPU in 
Eq.~(\ref{Eq2}) polynomially increases with the increasing number of $\phi$, even though the CPU can 
calculate significantly large sizes compared to a GPU.

\begin{figure}
\includegraphics[width=1.04\columnwidth]{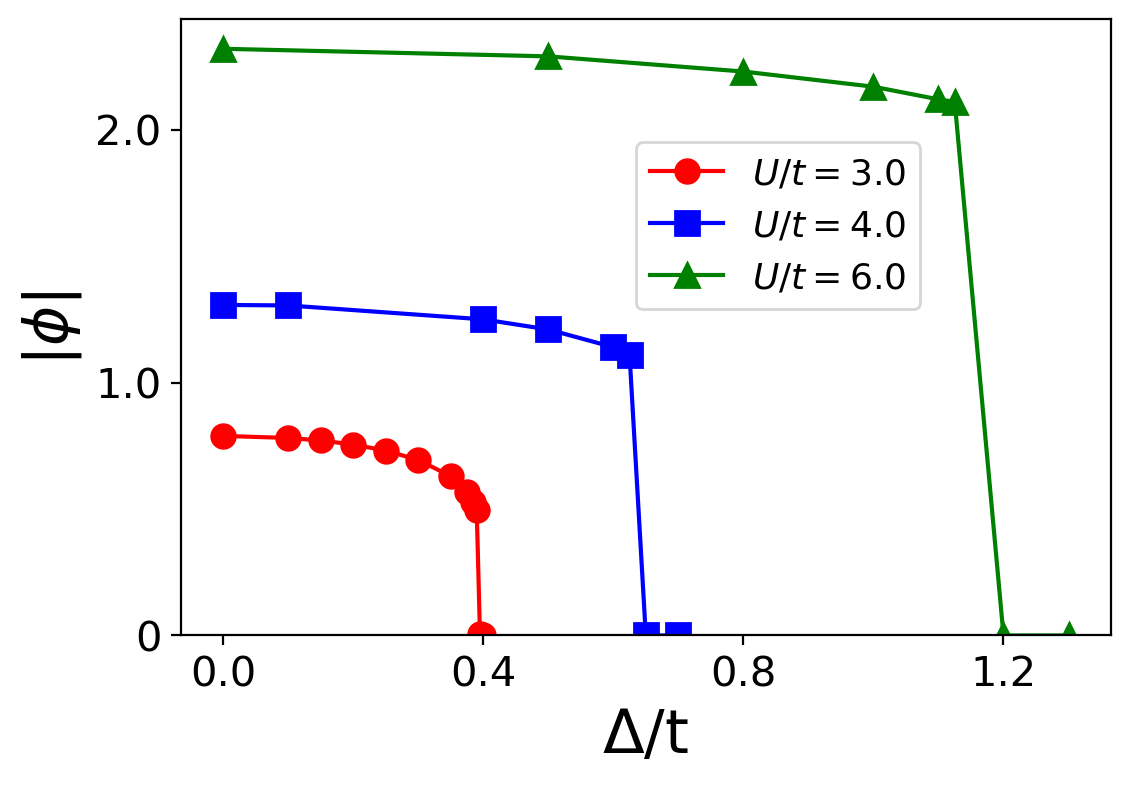}
\caption {\label{Fig5} (Color online) $\vert \phi \vert$ as a function of $\Delta/t$ for $U/t=2.0$, 
$4.0$, and $6.0$ in the half-filled ionic Hubbard model on 2D $20 \times 20$ square lattice
at $T/t=0.1$.}
\end{figure}

We evaluated $\rho_{\sigma} (\omega/t)$ and $\vert \phi \vert$ of the pure Hubbard model without $\Delta/
t$ for a half-filled 2D $L \times L$ square lattice with $L=20$. Figs.~\ref{Fig2} (a) and (b) show $\rho_{\sigma} 
(\omega/t)$ as a function of the real frequency $\omega/t$ at $U/t=6.0$ for $T/t=0.2$ and $1.3$. 
Evidently, 
the AF insulator with a gap exhibits a low $T/t=0.2$ in Figs.~\ref{Fig3}(a), because of the broken spin 
symmetry. The paramagnetic metal with van Hove peak at the Fermi level ($\omega/t=0.0$) are shown at 
a high $T/=1.3$ in Figs.~\ref{Fig3}(b). We also show $\vert \phi \vert$ as a function of $T/t$ for several 
$U/t$ values in Fig.~\ref{Fig4}. An AF insulator with a finite $\vert \phi \vert$ is shown at low $T$ for 
all $U/t$. The Neel $T/t$ that eliminates $\vert \phi \vert$ in the AF order increases with $U/t$. 
The result does not show the Mott insulator at high $T/t$ in strong interaction $U/t$ and is similar to 
that obtained using the Hartree-Fock approximation because the SCA 
ignores charge fluctuations. However, the SCA based results are 
slightly better than those obtained using the Hartree-Fock approximation, because dynamical fluctuations 
are considered at the zero Matsubara frequency level for the static imaginary time $\tau$ beyond the 
Hartree-Fock approximation.

\begin{figure}
\includegraphics[width=1.04\columnwidth]{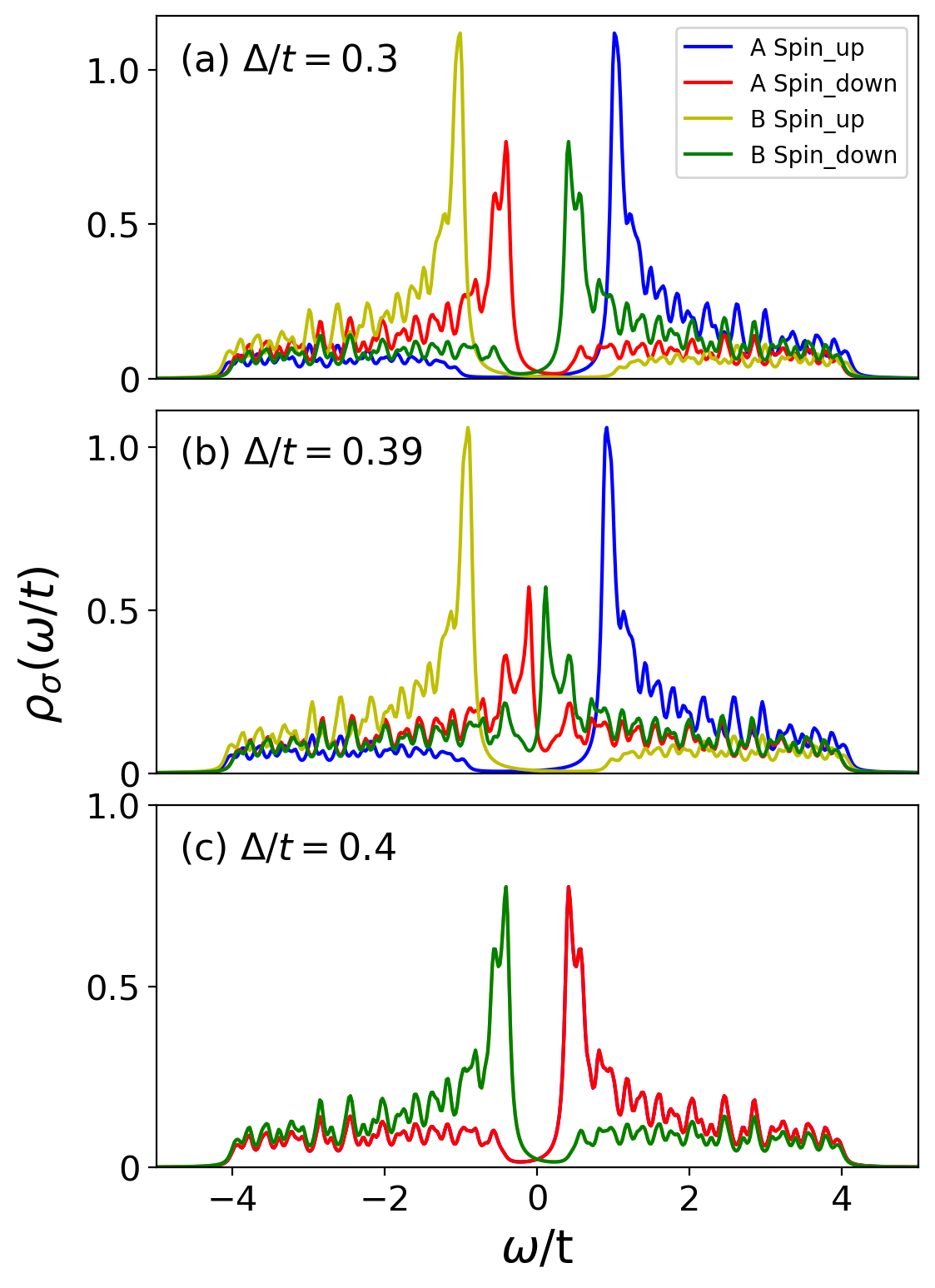}
\caption {\label{Fig6} (Color online) $\rho_{\sigma} (\omega/t)$ as a function of real
frequency $\omega/t$ in the ionic Hubbard model on 2D $20 \times 20$ square lattice at $U/t=3.0$ and $T/
t=0.1$ for (a) $\Delta/t=0.3$, (b) $0.39$ and (c) $0.4$.}
\end{figure}

Finally, we investigated the ionic Hubbard model, which displays a charge density wave (CDW), an AF 
insulator, and a metal. Fig.~\ref{Fig5} shows $\vert \phi \vert$ as a function of $\Delta/t$ for 
$U/t=2.0$, $4.0$, and $6.0$ in the half-filled ionic Hubbard model on 2D $20 \times 20$ square lattice
at $T/t=0.1$. Evidently, $\vert \phi \vert$ gradually decreases around the critical ionic 
potential $\Delta^*/t=0.39$ for $U/t=3.0$, whereas it shows an abrupt decrease at $\Delta^*/t=0.68$ and 
$1.19$ for $U/t=4.0$ and $5.0$, respectively. We estimate that the gently falling shape around $\Delta^*/
t=0.39$ of $U/t=3.0$ can be attributed to the presence of the metallic state between the CDW and AF 
insulators. As $U/t$ increases, the gently dropping curve suddenly shows an abrupt drop at $\Delta^*/t$. 
This change in the curve indicates the absence of metallic state between the CDW and AF insulators. 
In order to confirm the presence of metallic state in weak interaction region of $U/t=3.0$, 
we plotted $\rho_{\sigma} (\omega/t)$ at $U/t=3.0$ and $T/t=0.1$ for $\Delta/t=0.3$, $0.39$ and $0.4$ in 
Figs.~\ref{Fig6}(a), (b) and (c), respectively. A gap opening in $\rho_{\sigma} (\omega/t)$ is broken by 
the spin symmetry in weak $\Delta/t$ of Fig.~\ref{Fig6}(a), while it is induced by broken priodic 
potential in strong $\Delta/t$ of Fig.~\ref{Fig6}(c). Two different phases in Figs.~\ref{Fig6}(a) and (c) 
mean the AF insulator and CDW, respectively. We confirmed the AF metal with broken spin symmetry and 
finite density at Fermi level in intermediate region of Fig.~\ref{Fig6}(b).

\section{Conclusion\label{summary}}

The Hubbard model of a 2D $L \times L$ square lattice is an unsolved problem in physics. Exact numerical 
methods, such as ED and QMC, are limited by the size of the 2D lattice. Therefore, developing approximate 
methods to compute the physical properties of large-scale lattice sizes is necessary. In this study, we 
developed the SCA+ADAM approach to compute $\rho_{\sigma}(\omega/t)$ and $\vert \phi \vert$ of a 2D ionic 
Hubbard model with long-range spatial correlations within an appropriate computational duration. We 
evaluated the computation time as a function of the lattice size on the CPU and GPU as well as examined 
the physical properties of the Hubbard model without ionic potential and compare them with the results 
computed using the Hartree-Fock approximation method. Finally, the one-particle properties and AF order 
parameter of the ionic Hubbard model were analyzed.

Notably, the disordered Hubbard model, which is characterized by a random onsite potential and is more 
interesting and complex than the ionic Hubbard model, requires a large lattice to capture the competition 
between electron hopping, random onsite potentials, and Coulomb interactions~\cite{Anderson1958}. The 
SCA+ADAM approach can be used to compute the physical properties of the Hubbard model with competetion 
between long-range spatial AF correlations and onsite random potential. Therefore, we believe that the 
proposed method can be applied to the disordered Hubbard model and other such fundamental physical 
problems in the future.

\section{Acknowledgements}
This work was supported by Ministry of Science through NRF-2021R1111A2057259 funded by the Korean 
government. We would like to acknowledge the hospitality at APCTP where part of this work was done.

\end{document}